# Sensitivity of a low threshold directional detector to CNO-cycle solar neutrinos

R. Bonventre[1,a], G. D. Orebi Gann[1,2]

[1] Lawrence Berkeley National Laboratory, Berkeley, CA 94720-8153, USA
[2] University of California, Berkeley, CA 94720-7300, USA

**Abstract** A first measurement of neutrinos from the CNO fusion cycle in the Sun would allow a resolution to the current solar metallicity problem. Detection of these low-energy neutrinos requires a low-threshold detector, while discrimination from radioactive backgrounds in the region of interest is significantly enhanced via directional sensitivity. This combination can be achieved in a water-based liquid scintillator target, which offers enhanced energy resolution beyond a standard water Cherenkov detector. We study the sensitivity of such a detector to CNO neutrinos under various detector and background scenarios, and draw conclusions about the requirements for such a detector to successfully measure the CNO neutrino flux. A detector designed to measure CNO neutrinos could also achieve a few-percent measurement of pep neutrinos.

## 1 Introduction

### 1.1 Solar neutrinos

Solar neutrino experiments were pivotal in the groundbreaking discovery of neutrino oscillation and, hence, massive neutrinos, while at the same time confirming our understanding of fusion processes in the Sun. The Sudbury Neutrino Observatory (SNO) experiment resolved the so-called Solar Neutrino Problem by detecting the Sun's "missing" neutrinos, confirming the theory of neutrino flavor change. The combination of a charged-current (CC) measurement from SNO with SuperKamiokande's high-precision elastic scattering (ES) measurement demonstrated that the electron neutrinos produced in the Sun were transitioning to other flavors prior to detection [1], a result later confirmed at $5\sigma$ by SNO's measurement of the flavor-independent $^8$B flux using the neutral current (NC) interaction [2]. The KamLAND reactor experiment confirmed this flavour change as being due to oscillation [3]. This opened the door to a precision regime, allowing neutrinos to be used to probe the structure of the Sun, as well as the Earth and far-distant stars. Solar neutrinos remain the only sector of neutrinos with a confirmed observation of the effect of matter on neutrino oscillation at high significance, providing a unique opportunity to further probe this interaction to search for non-standard interactions and other effects.

The Borexino experiment made the first direct measurements of the $^7$Be, *pp* and *pep* fluxes [4–6], as a result of which the *pp* fusion cycle in our Sun has been well studied, with measurements of all neutrino sources bar the high-energy *hep* neutrinos. The subdominant CNO cycle is less well understood and yet has the potential to shed light on remaining mysteries within the Sun. One of the critical factors that engendered confidence in the Standard Solar Model [7] (SSM) was the excellent agreement ($\sim 0.1\%$) of SSM predictions for the speed of sound with helioseismological measurements. The speed of sound predicted by the SSM is highly dependent on solar dynamics and opacity, which are affected by the Sun's composition [8]. In recent years the theoretical prediction for the abundance of metals (elements heavier than H or He) in the photosphere has fallen due to improvements in the modeling of the solar atmosphere, including replacing previous one-dimensional models with fully three-dimensional modeling, and inclusion of effects such as stratification and inhomogeneities [9]. The new results are more consistent with neighboring stars of similar type, and yield improved agreement with absorption-line shapes [10], but at the same time reduce the prediction for the metal abundance by $\sim 30\%$. When these new values are input to the SSM, the result is a discrepancy in the speed of sound with helioseismological observations. This new disagreement has become known as the "Solar Metallicity Problem" [11,12]. A measurement of the CNO neutrino flux may help in resolving this problem [13,14]. The impact of metal-licity on pp-chain neutrinos is small relative to theoretical uncertainties, but the neutrino flux from the sub-dominant CNO cycle depends linearly on the metallicity of the solar core, and the predictions for the two models differ by greater than 30% [15]. The theoretical uncertainty on these predictions is roughly 14–18% in the so-called AGS05-Opt model, although greater in other models [15]. However, these uncertainties can be reduced to < 10% using correlations in the theoretical uncertainties between the CNO and $^8$B neutrino fluxes: the two have similar dependence on environmental factors, thus a precision measurement of the $^8$B neutrino flux can be used to "calibrate" the core temperature of the Sun and, thus, constrain the CNO neutrino flux prediction [13]. In [13], the final uncertainty is dominated by the nuclear physics. A precision measurement of the CNO flux then has the potential to resolve the current uncertainty in heavy element abundance.

[a] e-mail: rbonventre@lbl.gov

Borexino has placed the most stringent limits on the CNO neutrino flux to date [6], and continues to pursue a first observation. However, extraction of this flux is extremely challenging due to the similarity of the spectrum of ES recoil electrons with background $^{210}$Bi decays in the target. Borexino propose to use the time evolution of the $\alpha$ decay of the daughter, $^{210}$Po, to constrain the level of $^{210}$Bi. This method requires both a stable $\alpha$-particle detection efficiency and a lack of external sources of the $^{210}$Po daughter, which can be challenging to achieve [16]. A recent paper discusses the sensitivity of several current and future experiments to the CNO flux [17]. A detector with directional sensitivity could discriminate between the directional solar neutrino signal and the isotropic background, without the need for a time-series analysis.

### 1.2 Low-threshold directional detection

Water Cherenkov detectors (WCD) are limited in energy threshold by the relatively low Cherenkov photon yield. Scintillator-based detectors can achieve the thresholds required to observe CNO neutrinos, but lose the advantage of the directional information provided by Cherenkov light. The novel water-based liquid scintillator (WbLS) target medium [18] offers the potential to benefit from both the abundant scintillation and directional Cherenkov signals, thus achieving a massive, low-threshold directional detector.

The sensitivity of a 50 kT pure LS detector has been studied by the LENA collaboration [19]. While the threshold of a WbLS detector will not be as low as for a pure LS target (LENA studies assumed a 250 keV threshold), the additional information provided by the Cherenkov component provides a strong benefit in signal/background separation. WbLS offers a uniquely broad, multi-parameter phase space that can be optimized to maximize sensitivity to a particular physics goal. The WbLS "cocktail" can range from a high-LS fraction, oil-like mixture, with > 90% LS, to a water-like mixture with anything from 10% to sub-percent levels of LS. The choice of fluor affects the scintillation yield, timing, and emission spectrum, and metallic isotopes can be deployed to provide additional targets for neutrino interaction [20].

In this article we study the sensitivity of a large WbLS detector to CNO neutrinos under a range of detector scenarios, including target size, LS fraction, photo-coverage, angular resolution, and the level of intrinsic radioactive contaminants. These studies can inform the design of a future experiment targeting a CNO flux observation, such as THEIA [21,22].

There is much interest in the community in developing low-threshold directional detectors. Monte Carlo studies in [23,24] discuss how the potential for separation of a Cherenkov signal in a scintillating target could be used to extract particle direction. The CHESS experiment has recently demonstrated first detection of a Cherenkov signal in pure LS (both LAB and LAB/PPO) [25,26]. While high-energy muons were used for this demonstration, they were in the MIP regime and thus the energy deposited along the few-cm track in the CHESS target was only a few MeV, within the regime relevant for this work. Studies based on data from the KamLAND detector show the potential for directional reconstruction using time-of-flight of the isotropic scintillation light [27,28].

In Sect. 2 we describe the analysis method for evaluating the uncertainty on the CNO and *pep* neutrino fluxes, and the simulation of each signal and background source. In Sect. 3 we describe various scenarios for both detector configuration and background assumptions under which the CNO flux is evaluated. Section 4 presents the results, and Sect. 5 describes the conclusions.

## 2 Analysis methods

Neutrino flux sensitivities are determined using a binned maximum likelihood fit over two-dimensional PDFs in energy and direction relative to the Sun, $\cos\theta_\odot$. The energy dimension allows separation of neutrino fluxes from each other, as well as discrimination from certain background events. The direction dimension is critical for a full separation of the CNO flux from radioactive background.

### 2.1 Simulation of expected signals

Simulations for this paper were produced using the RAT-PAC software (https://github.com/rat-pac/rat-pac), which is based on Geant4. Optical properties for the WbLS were constructed from weighted combinations of water and scintillator optics as determined by the SNO+ collaboration. The optical simulation was tuned to allow multicomponent absorption and reemission with separate absorption lengths and reemis-sion probabilities for each component of the WbLS cocktail. Radioactive decays were simulated using the decay-chain generator developed by Joe Formaggio and Jason Detwiler (Private communication), and solar neutrino interactions were simulated using an elastic scattering generator also developed by Joe Formaggio. Solar signals were simulated assuming fluxes from the BS05OP solar model [7], which assumes the higher solar metallicity, and using LMA-MSW survival probabilities from the three flavor best fit oscillation values from [29].

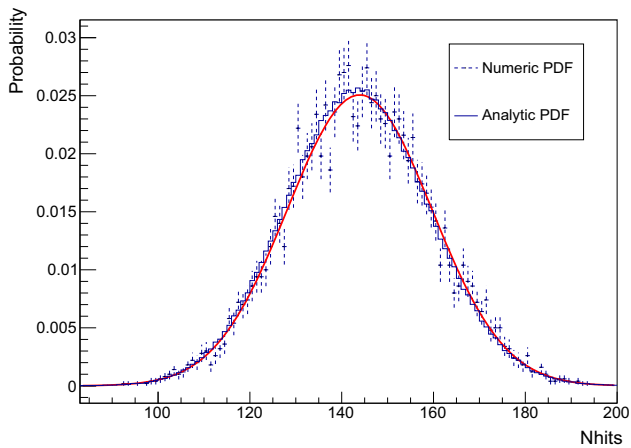

**Fig. 1** Comparison of NHit distribution for 1 MeV electrons isotropically distributed in a 50 kT detector simulated fully with Geant4 (numeric PDF) and with only Cherenkov light simulated and scintillation light contribution determined from pregenerated tables (analytic PDF)

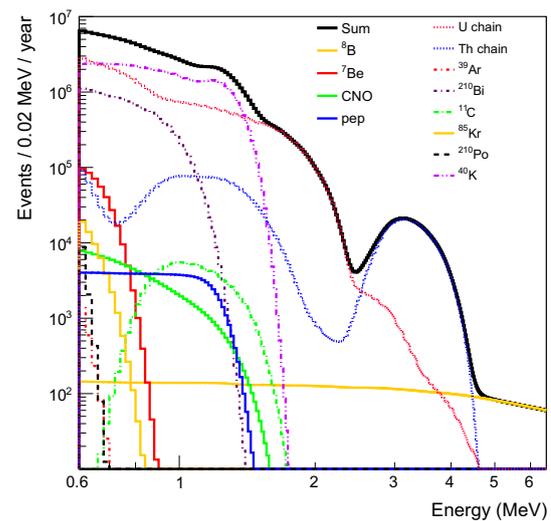

**Fig. 2** Expected energy spectra for the baseline detector configuration and background assumptions, within a 50% fiducial volume

The reconstructed effective electron energy spectrum for each signal was determined semi-analytically. First, the distribution of the number of PMT hits (NHit) per event for each signal was found, handling the scintillation and Cherenkov components separately. Given a WbLS cocktail and detector configuration, the scintillation contribution to the NHit for an event at a specific position scales linearly with the quenched energy deposition. This scaling was determined by simulating electrons at each position with the Cherenkov light production disabled. The Cherenkov contribution to the NHit was determined by simulating each signal with the scintillation light yield set to zero, but with absorption and reemission of the Cherenkov light in the scintillator and wavelength shifter enabled. The expected NHit for each event was taken to be the sum of the Cherenkov Nhit plus a number of scintillation hits drawn from a Poisson distribution with a mean given by the scintillator energy deposition in that event times the scaling factor for the relevant event position. This method made efficient use of computing resources to simulate a full set of background event types. The result of this procedure was compared with the NHit distribution from a full simulation for 1 MeV electrons, and both the mean and width were found to agree to within 0.5%, as shown in Fig. 1.

The conversion from NHit to reconstructed effective electron energy was determined using a position- and direction-dependent lookup table. The lookup table was generated by simulating electrons at various positions, directions, and energies using the above procedure. The resultant energy PDFs for 5% scintillator are shown in Fig. 2.

For each signal the $\cos\theta_\odot$ distribution was determined fully analytically. All non-neutrino signals were assumed to be flat. For the solar signals the ES generator determined the electron direction relative to the Sun based on the differential cross sections. This was then convolved with a chosen angular resolution (Sect. 3).

2.2 Baseline fit configuration

The normalizations for 11 signals are floated in the fit. $^{238}$U and $^{234}$Th chain backgrounds are assumed to be in equilibrium except for $^{210}$Bi, $^{210}$Po, and $^{210}$Pb, and the backgrounds in each chain are floated together as a single parameter. The various flavor components of the $^{8}$B, $^{7}$Be, pep, and CNO signals are combined into one parameter per flux, assuming survival probabilities from [29]. The CNO signal contains the sum of the $^{17}$F, $^{15}$O, and $^{13}$N solar neutrino signals. The *pp* solar signal is not included as it falls below the energy threshold, and the *hep* solar signal is fixed in magnitude as it is too small to be reconstructed. $^{210}$Bi, $^{40}$K, $^{85}$Kr, and cosmogenically activated $^{11}$C are each included as a separate parameter. The $^{39}$Ar and $^{210}$Po backgrounds are floated together as their energy spectra above 600 keV are similar.

The baseline fit uses 40 bins in $\cos\theta_\odot$ and 20 keV bins in energy from 600 keV to 6.5 MeV. A 5 year livetime is assumed.

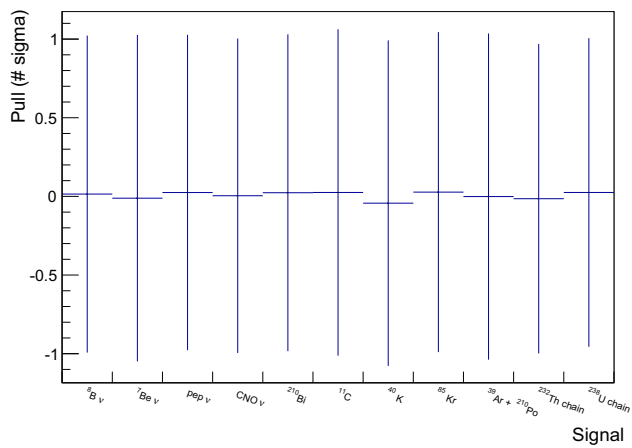

**Fig. 3** Pull distribution for fits to the baseline detector configuration and background assumptions. Note that the error bars show the RMS, and not the error on the mean. A mean pull of 0 with an RMS of 1 is expected for each signal in an unbiased fit.

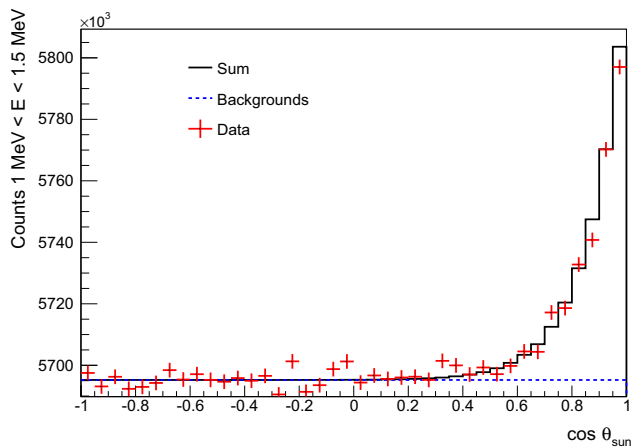

**Fig. 4** Projection of two dimensional fit in $\cos\theta_\odot$ for energies between 1 and 1.5 MeV for a randomly generated fake dataset with the baseline detector configuration and background assumptions

A 50% fiducial volume cut is applied in order to reduce the impact of background contaminants in external regions of the detector, such as $\gamma$s from $^{208}$Tl in the PMTs, to negligible.

Full bias and pull studies were performed, and the fit was observed to be unbiased, with the expected pull distribution. Figure 3 shows the pull distribution for fits with the baseline configuration.

Figure 4 shows the projection of one two-dimensional fit in the $\cos\theta_\odot$ dimension. This figure illustrates the importance of angular resolution in extracting the solar neutrino signal even with backgrounds many order of magnitudes larger, using the high statistics achievable in a large detector.

## 3 Detector configuration and background assumptions

The sensitivity to CNO neutrinos is studied under a range of detector scenarios and background assumptions:

1. Target volume.
2. Angular resolution.
3. WbLS cocktail.
4. Photocathode coverage.
5. Energy resolution.
6. Energy threshold.
7. Background assumptions.

For the purposes of comparison we define the baseline detector configuration to be a 50-kT detector with 90% PMT coverage, 5% WbLS, and 25° angular resolution, with baseline background levels as given in Table 2 in Sect. 3. All results assume a five year livetime.

*Target volume* We consider both a 25 and 50 kT total detector volume, corresponding to a 31.7 or 40-m sized right-cylindrical vessel. The PMTs are positioned at the edge of this volume, and a fiducial volume is selected for analysis (Sect. 2). The 50% fiducial volume corresponds to a 4.15-m buffer between the PMTs and the target volume for the 50 kT detector, and a 3.27-m buffer for the 25 kT detector.

Increasing the target volume scales the exposure accordingly, but at the same time reduces the overall light collection of the detector due to absorption, thus impacting the achievable energy resolution. This additional absorption would also negatively impact the angular resolution. Since this work does not perform a full directional reconstruction, this correlation is not explicitly included. However, a range of possible angular resolutions are considered for each target volume.

*Angular resolution* A critical factor in this work is the assumed angular resolution. The angle between the incoming particle direction and the direction to the Sun, $\cos\theta_\odot$, can be used to differentiate signal from background. Due to the kinematics of the ES interaction, solar neutrinos are predominantly directed away from the Sun. This provides a key handle to discriminate solar neutrino events from an isotropic radioactive background.

All radioactive backgrounds are assumed to be isotropically distributed and thus have a flat distribution in $\cos\theta_\odot$. For solar neutrinos the direction of the electron relative to the Sun was determined as a function of energy via simulations that take as input the full differential cross sections. The resulting electron direction was then convolved with a detector angular resolution of the form

$$e^{\frac{1}{\sigma}(\cos\theta - 1)}, \qquad (1)$$

with a width given by $\sigma$. The angular resolution was assumed to be constant with energy, with the value defined at threshold. Any improvement in sensitivity as the angular resolution improves at energies significantly above the energy threshold was observed to be a second order effect.

This work does not attempt a full reconstruction of event direction. Instead, we consider a range of possible angular resolutions in order to determine the impact on the final neutrino flux sensitivity. In the best case, we consider a resolution of 25° – similar to that achieved by SNO. In SNO, a 1-kT heavy-water detector with 55% PMT coverage, an angular resolution of 26.7° was achieved for $^{16}$N events (approximately 5 MeV), which had an average of 36 PMT hits [30]. In Super Kamiokande, a 50-kT light-water detector with 40% coverage, an angular resolution of approximately 35° was achieved at 6 MeV, with 41 PMT hits [31]. Table 1 shows the expected number of hits from Cherenkov photons in the proposed 0.5% WbLS detector for a range of energy thresholds. As shown, by 1 MeV we expect as many hits from Cherenkov photons as in SNO and Super Kamiokande at 5 or 6 MeV. While many of these photons may be scattered, or absorbed and reemitted by the scintillator, absorption lengths in WbLS are significantly longer than a pure LS detector. Additionally, some directional information is retained by considering the offset of the reemitted photon from the original production point. The proposed detector has several further advantages that could allow for greatly improved angular resolution at lower energies. Increased coverage and use of high quantum efficiency PMTs allows detection of many more direct Cherenkov photons than in SNO and Super Kamiokande at equivalent energies. The inclusion of wavelength shifter in the WbLS absorbs and reemits Cherenkov photons that would otherwise be at too small a wavelength to be detected by the PMTs and, depending on the absorption length, may also retain some directional information. As demonstrated in [27,28], even a pure LS detector can provide some directional information by considering photon time of flight. With these advantages, a resolution of 25° may be achievable.

As the detector size increases, coverage is reduced, or the LS fraction is increased, the resolution will naturally be degraded by increased scattering and absorption, and reduced light collection. We find that a large fraction of wavelength shifted photons are absorbed within a short distance, and so it is possible a lower concentration of PPO would be desirable to better retain directional information. To estimate the impact of these effects we consider degraded resolutions of 35°, 45°, and 55° for each detector scenario.

*WbLS cocktail* The scintillator component of the WbLS cocktail is taken to be LAB with 2g/L of PPO as a fluor (hereafter referred to as LAB/PPO). LAB/PPO properties have been determined by the SNO+ collaboration [32]. We consider fractions of LS from 0.5 to 5%, as well as a pure LS detector. Changing the fraction of LS in the cocktail affects the overall scintillation light yield and, thus, energy resolution. This effect is incorporated into the fit via the energy reconstruction described in Sect. 2.1. The energy resolution achieved in a 50 kT detector with 90% coverage for various WbLS fractions is shown in Fig. 5.

**Table 1** Expected number of PMT hits from Cherenkov photons for a 90% PMT coverage detector. The first two columns include scattered photons and those absorbed and reemitted by wavelength shifter, while the second two columns include only photons that make it to the PMT directly without scattering or absorption. Results are shown across the range of energy thresholds considered for the analysis (Sect. 3). For comparison, assuming 1 kHz dark rates in the PMTs, we expect 1.2 PMT hits in the 25 kT detector and 1.9 hits in the 50 kT detector from noise within a prompt 20 ns time window

|       | All Cherenkov photons | | Direct photons | |
|-------|---------|---------|---------|---------|
|       | 0.6 MeV | 1.0 MeV | 0.6 MeV | 1.0 MeV |
| 25 kT | 12.5    | 37.6    | 3.3     | 10.4    |
| 50 kT | 11.5    | 34.7    | 3.0     | 9.5     |

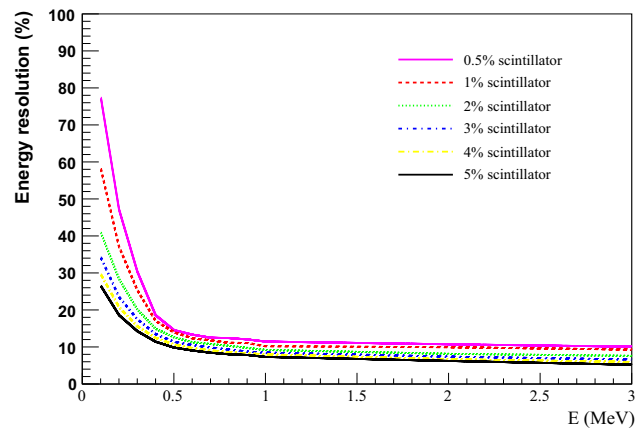

**Fig. 5** Energy resolution as a function of energy for various scintillator fractions for a 50 kT detector

The Cherenkov light yield is unaffected at first order by this change to the target cocktail, although there is a non-zero impact through absorption and reemission in the LS, which is fully modeled in the simulation. While this change can therefore be expected to affect the angular resolution, this effect is not included in the PDFs since this work does not perform a full directional reconstruction. Instead, we consider the impact of a range of angular resolutions for each target cocktail.

*Photocathode coverage* We assume instrumentation with Hamamatsu R11780 12 inch high quantum efficiency (HQE) PMTs, which have a peak efficiency of 32% at 390 nm [33]. We study photocathode coverages of up to 90%, which would require approximately 100k PMTs for the 50 kT detector. Changing the photocathode coverage effectively scales the number of detected PMT hits per MeV of deposited energy, thus scaling both Cherenkov and scintillation contributions. Figure 6 shows the PMT QE, and the emission spectra for both Cherenkov and scintillation light.

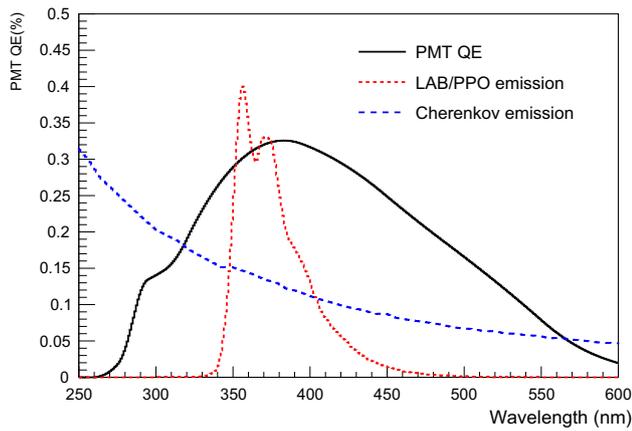

**Fig. 6** Shape of the QE curve used for Hamamatsu R11780 HQE PMTs along with the Cherenkov and scintillation emission spectra. The normalization for the emission spectra is arbitrary

*Energy scale and resolution* The energy resolution is determined by the overall light yield, which is dependent on the WbLS cocktail and the photocathode coverage, the impact of each of which is studied separately (Sect. 3). Here, we consider the impact of systematic uncertainties in the energy scale and resolution on the analysis. We investigate the effect of uncertainties in energy by modifying the PDFs relative to the spectra from which fake datasets are drawn. For the energy scale, a linear shift in energy is applied to the energy spectra. For the resolution a Gaussian smearing is applied to the energy spectra, with the width of the Gaussian set to a constant value.

*Energy threshold* The ability to reach a lower energy threshold than a water detector is one of the main advantages of a WbLS detector besides increased energy resolution. We consider energy thresholds from 600 keV up to 1 MeV. The possibility of reaching thresholds this low depend on the background rate, the PMT dark noise rate, the trigger setup, and the sustainable data rate. At 0.5% scintillator, we expect a total of 19.3 PMT hits at 0.6 MeV in a 50 kT detector, and by 5% scintillator we expect 93.0, compared to 5.4 PMT hits in a water only detector. Assuming a trigger window of 200 ns as used in Super Kamiokande, with ∼ 94,000 PMTs, we expect 18.8 noise hits per trigger window per kHz PMT dark rate, and so at 0.5% scintillator a dark rate lower than 1 kHz would be required. In a 25 kT detector we expect 21.1 PMT hits at 0.6 MeV, and 11.7 noise hits per trigger window per kHz PMT dark rate.

**Table 2** Background assumptions for the baseline configuration

|  | $H_2O$ level (g/g$H_2O$) | LS Level (g/gLAB) |
|---|---|---|
| $^{238}$U chain | 6.63e−15 [37] | 1.6e−17 [35] |
| $^{232}$Th chain | 8.8e−16 [37] | 6.8e−18 [35] |
| $^{40}$K | 6.1e−16[a] | 1.3e−18 [36] |
| $^{85}$Kr | 2.4e−25[b] | 2.4e−25 [36] |
| $^{39}$Ar | 2.75e−24[b] | 2.75e−24 [36] |
| $^{210}$Bi | 3.78e−28[b] | 3.78e−28 [36] |
| $^{11}$C | 0 | 1.0e5 (ev/kT/year) [4] |

[a]The $^{40}$K level in water is taken to be 0.1× the Borexino measurement [34]
[b]The $^{85}$Kr, $^{39}$Ar, and $^{210}$Bi levels in water are taken to be the Borexino measured level in scintillator [36], although levels increased by several orders of magnitude are explored

*Background assumptions* Radioactive contaminants in the target material are calculated as the sum of contamination from the LS and the water components, weighted by the corresponding mass fractions of the WbLS cocktail. Table 2 details the numbers assumed in the baseline analysis, taken from measurements by SNO and Borexino [34–37]. In the analysis we consider a range of levels for the intrinsic contamination in the WbLS target, as well as the degree of $\alpha$–$\beta$ separation and Bi-Po pile-up rejection achievable.

- U- and Th-chain: The LS components of uranium- and thorium-chain background levels are assumed to be at Borexino levels and the water components are assumed to be at the level of the heavy water in SNO [35,37]. The baseline alpha rejection is assumed to be 95%. For $^{212}$Bi-$^{212}$Po and $^{214}$Bi-$^{214}$Po events, an event window of 400 ns is assumed, with a 95% rejection for in-window coincident events and 100% rejection for tagging out of window events. While this level of discrimination has not yet been demonstrated in a WbLS target, this is a future goal of the CHESS experiment [25]. The discrimination achieved will depend on both the $\alpha$ quenching and overall light yield of the target, as well as specific timing properties. These microphysical properties must be fully understood in order to quantify the level of $\alpha$ and Bi-Po coincidence rejection that can be achieved. The impact of the efficiency of both $\alpha$ rejection and Bi-Po tagging is studied.
- $^{40}$K: The level of $^{40}$K in LS is a conservative estimate from the upper limit of Borexino's initial measurement. The level in water is taken from an upper limit measured in the Borexino Counting Test Facility (CTF) [34]. SNO measured a level of 2e−9 gK/g$H_2O$ in the light water, although this background was below threshold in SNO and thus little effort was made to reduce it [38]. These measurements are therefore taken as conservative upper bounds on the level. We use $0.1\times$ the Borexino level in water as the baseline level for this study, and investigate the impact of a contamination an order of magnitude higher.

- $^{210}$Bi: $^{210}$Pb, $^{210}$Bi, and $^{210}$Po cannot be assumed to be in equilibrium with the rest of the uranium chain due to the long half life of $^{210}$Pb and the possibility of Rn contamination. The baseline level of $^{210}$Bi in LS is taken from Borexino. The level of out-of-equilibrium background achievable in ultra-pure water has not been measured to the precision needed for this kind of experiment. The uranium-chain contribution from the water component is orders of magnitude larger than the $^{210}$Bi level measured in Borexino, thus the contribution from any out-of-equilibrium component in water must be many orders of magnitude larger than in LS in order to impact the sensitivity. Values of $10\times$, $100\times$ and $1000\times$ the Borexino-measured value in scintillator are explored for the contamination in water.

- $^{85}$Kr and $^{39}$Ar: The levels of $^{85}$Kr and $^{39}$Ar in LS are taken from Borexino. The levels in water are not well known; various multiples of the LS level are explored in the analysis, up to $10{,}000\times$ the Borexino level.

- $^{11}$C: A potential site for the detector is the LBNF site, which provides an overburden of 4850 feet (4300 m.w.e.) to shield from cosmogenic backgrounds. The $^{11}$C level in Borexino, with 3800 m.w.e. overburden, is approximately 1e5 events per kiloton year. This is used as a conservative initial estimate for the rate, adjusted for the carbon content of the different target materials. Rates of an order of magnitude higher and lower are studied, to simulate the effect of different possible detector sites. Production of cosmogenic backgrounds on the water component of the target were considered according to [39], which provides a complete list of potential spallation products. The dominant sources in the energy range considered in this work are $^{11}$C and $^{15}$O. Inclusion of these backgrounds was observed to change the CNO sensitivity by 0.03% for the baseline configuration, and the pep sensitivity by an unobservable amount. Thus, these backgrounds were omitted for the remainder of this work.

- Externals: Background contributions from radioactive contamination external to the target region (for example $^{208}$Tl $\gamma$s from the PMTs and any support structures) are assumed to be negligible inside a chosen 50% fiducial volume. In future studies, vertex reconstruction could be used to constrain such sources and thus potentially expand the fiducial volume.

**Table 3** Fit uncertainty for 5 years of data with the baseline configuration and background assumptions

| Signal | Normalization sensitivity (%) |
| --- | --- |
| $^8$B $\nu$ | 0.4 |
| $^7$Be $\nu$ | 0.4 |
| pep $\nu$ | 3.8 |
| CNO $\nu$ | 5.3 |
| $^{210}$Bi | 0.1 |
| $^{11}$C | 11.5 |
| $^{85}$Kr | 10.5 |
| $^{40}$K | 0.04 |
| $^{39}$Ar/$^{210}$Po | 21.9 |
| $^{238}$U chain | 0.02 |
| $^{232}$Th chain | 0.05 |

**Table 4** CNO flux sensitivity (%) as a function of target mass, WbLS % and angular resolution for 5 years of data with 90% PMT coverage and the baseline background assumptions

| Target mass | WbLS | Angular resolution | | | |
| --- | --- | --- | --- | --- | --- |
| | | 25° | 35° | 45° | 55° |
| 50 kT | 0.5% | 6.2 | 8.8 | 11.2 | 13.5 |
| 50 kT | 1% | 6.1 | 8.7 | 11.0 | 13.4 |
| 50 kT | 2% | 6.2 | 8.9 | 11.4 | 13.8 |
| 50 kT | 3% | 5.9 | 8.4 | 10.7 | 13.0 |
| 50 kT | 4% | 5.5 | 7.9 | 10.1 | 12.3 |
| 50 kT | 5% | 5.3 | 7.6 | 9.7 | 11.8 |
| 25 kT | 0.5% | 8.5 | 12.2 | 15.6 | 18.7 |
| 25 kT | 1% | 8.5 | 12.1 | 15.0 | 18.4 |
| 25 kT | 2% | 8.5 | 12.1 | 15.5 | 18.7 |
| 25 kT | 3% | 8.0 | 11.5 | 14.6 | 17.7 |
| 25 kT | 4% | 7.6 | 10.9 | 13.9 | 16.8 |
| 25 kT | 5% | 7.3 | 10.5 | 13.3 | 16.2 |

## 4 Results

The fit uncertainty for each signal with the baseline detector configuration and background assumptions is shown in Table 3.

*Detector size, target, and angular resolution* The CNO solar neutrino sensitivity as a function of detector size, LS fraction, and angular resolution for the baseline background assumptions is shown in Table 4 and Fig. 7.

As a comparison we look at simulated spectra from a 50 kT pure LS detector and study the results of a one dimensional fit in energy (under the assumption that there would be no directional resolution in pure LS). Here we find a CNO sensitivity of 3.5%, but pull distributions show that the fit is not able to converge on the full errors in the CNO and $^{210}$Bi signals (Fig. 8). The correlation between the two values in the fit is −0.84. This suggests than energy alone is not sufficient to distinguish these signals even at very high statistics.

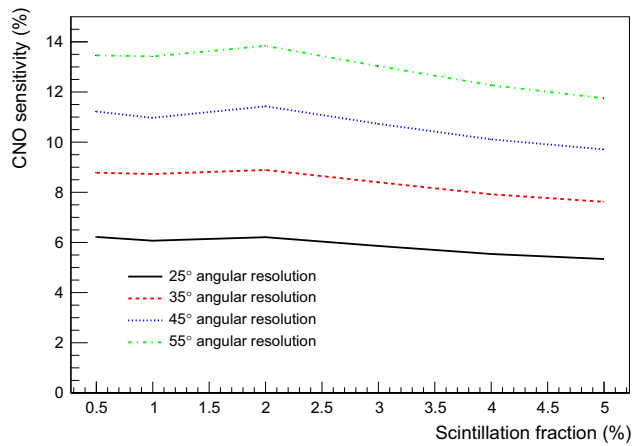

**Fig. 7** CNO sensitivity as a function of scintillator fraction and angular resolution for a 50 kT detector after 5 years of running with the baseline background assumptions

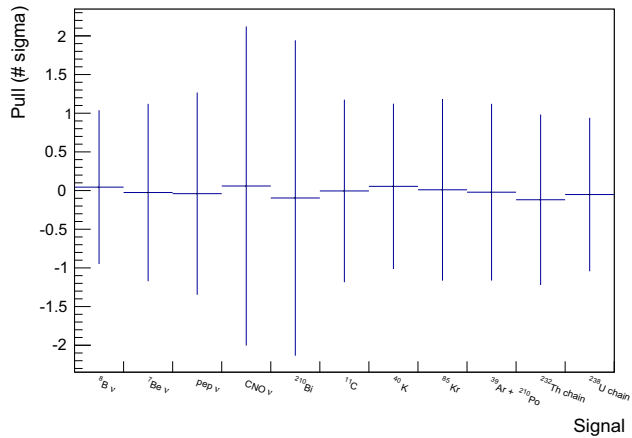

**Fig. 8** Pull distribution for 5 years of data with a 50 kT pure scintillator detector and the baseline background assumptions. Note that the error bars show the RMS, and not the error on the mean. A mean pull of 0 with an RMS of 1 is expected for each signal in an unbiased fit

*PMT coverage* The above results show that the impact of the improved energy resolution due to a larger scintillator fraction (0.5–5%) is marginal. However, changes in PMT coverage can have a potentially greater effect. At low energies a significant fraction of the resolution comes from Cherenkov photons, which does not scale with scintillator fraction. Reducing the PMT coverage will reduce the angular resolution as well as the overall light collection. The fit uncertainty for CNO for 5 years of data using a detector with 60% PMT coverage instead of 90% is shown in Table 5. We can see as predicted that at low scintillator fractions the change in PMT coverage has a larger effect than an equivalent fractional change in scintillator fraction. At higher scintillator fractions the effect is smaller, and suggests that the main consideration for the PMT coverage requirement will be the achievable angular resolution.

**Table 5** CNO flux sensitivity (%) as a function of target mass, WbLS % and angular resolution for 5 years of data and 60% PMT coverage with the baseline background assumptions

| Target mass | WbLS | Angular resolution | | | |
|---|---|---|---|---|---|
| | | 25° | 35° | 45° | 55° |
| 50 kT | 0.5% | 6.7 | 9.1 | 11.7 | 14.4 |
| 50 kT | 1% | 6.5 | 9.3 | 11.6 | 14.1 |
| 50 kT | 2% | 6.8 | 9.8 | 12.4 | 15.2 |
| 50 kT | 3% | 6.6 | 9.4 | 12.0 | 14.6 |
| 50 kT | 4% | 6.2 | 8.8 | 11.3 | 13.7 |
| 50 kT | 5% | 5.9 | 8.5 | 10.8 | 13.0 |
| 25 kT | 0.5% | 9.1 | 12.8 | 16.2 | 19.2 |
| 25 kT | 1% | 8.9 | 12.7 | 16.1 | 19.1 |
| 25 kT | 2% | 9.3 | 13.3 | 16.9 | 20.4 |
| 25 kT | 3% | 8.9 | 12.7 | 16.2 | 19.7 |
| 25 kT | 4% | 8.4 | 12.0 | 15.3 | 18.6 |
| 25 kT | 5% | 8.0 | 11.5 | 14.6 | 17.8 |

*Energy scale and resolution systematic uncertainties* We shift the PDFs in energy by a fixed percentage to determine the effect of uncertainties in the energy scale. The likelihood space is scanned to determine where this uncertainty can be constrained with the data by evaluating the change in negative log likelihood ($\Delta$NLL). Many fake data sets were studied, thus giving a range of results for the $\Delta$NLL. This range is represented with vertical error bars in the resulting figures (Figs. 9, 10).

For the 50 kT detector at 5% WbLS and 25° angular resolution the energy scale uncertainty can be constrained to less than 0.006%, where the change in the fitted CNO normalization is 1.7% (as shown in Fig. 9).

By smearing the PDFs with a Gaussian of pre-defined width, we investigate the impact of an uncertainty on the energy resolution. A scan of the likelihood space for the 50 kT detector at 5% WbLS and 25° angular resolution demonstrates the capability to constrain this uncertainty with the data itself down to 5.5 keV, at which point the systematic change in the fitted CNO normalization is 3.6%, as shown in Fig. 10.

*Energy threshold* The CNO solar neutrino sensitivity as a function of the energy threshold for the likelihood fit is shown in Table 6. These results show that a precise measurement of the CNO flux requires maintaining sensitivity below 1 MeV. It can be seen in Fig. 2 that this energy region is where the CNO and pep solar neutrino signals can be distinguished from each other, and at higher energy thresholds these signals become highly correlated.

*Background assumptions* As seen in Fig. 2, $^{40}$K is a dominant background at low energies, even assuming an order of magnitude improvement over the level measured in water by Borexino and SNO. This is due to the much higher contamination in the water component of the WbLS compared to the relatively cleaner scintillator. The $^{40}$K background in water was not critical for these previous measurements, and so it may be possible to further reduce the level with additional effort. The SNO water processing plant could be improved by increasing the frequency of replacing ion exchange columns or by distilling the water. Table 7 shows that at the Borexino measured level this background greatly decreases the sensitivity in all configurations compared to the baseline background assumptions which includes the order of magnitude reduction. Table 8 shows the sensitivity to each floated signal in the fit for the baseline detector configuration.

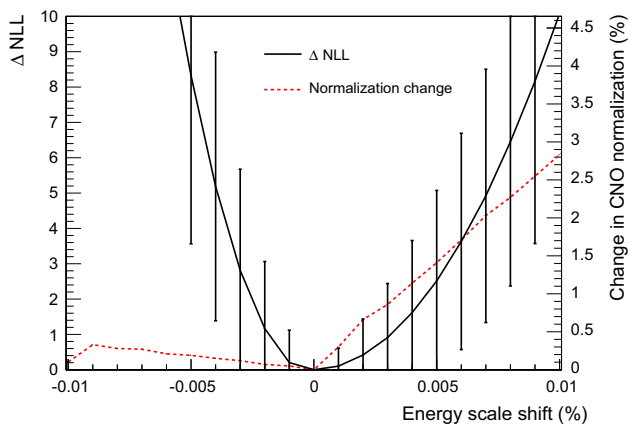

**Fig. 9** Result of systematic change in energy scale modeled by a fixed percentage shift in the energy PDF. The solid line shows the $\Delta$NLL from fitting with a shifted PDF relative to the true PDF, with the error bars showing the RMS of the $\Delta$NLL for many fake data sets. The dashed line shows the change in the mean extracted normalization of the CNO flux when fitting with a shifted PDF. Results are for 5 years of data with the baseline detector configuration and background assumptions

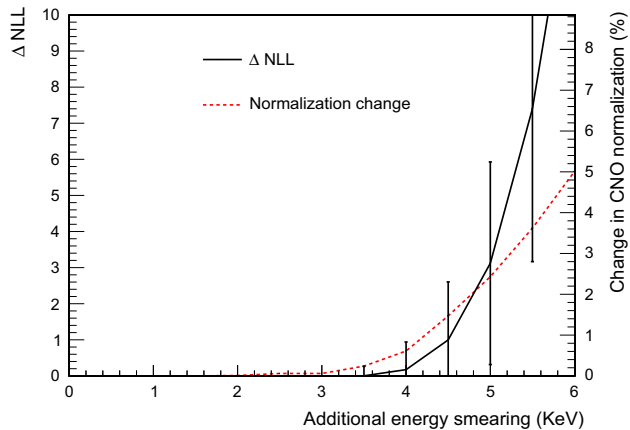

**Fig. 10** Result of systematic change in energy resolution modeled by an additional fixed width smearing of the energy PDF. The solid line shows the $\Delta$NLL from fitting with a shifted PDF relative to the true PDF, with the error bars showing the RMS of the $\Delta$NLL for many fake data sets, while the dashed line shows the change in the mean extracted normalization of the CNO flux when fitting with a shifted PDF. Results are for 5 years of data with the baseline detector configuration and background assumptions

**Table 6** Fit uncertainty for 5 years of data with the baseline detector configuration as a function of the energy threshold

| Energy threshold (MeV) | CNO sensitivity (%) |
|---|---|
| 0.6 | 5.3 |
| 0.7 | 5.9 |
| 0.8 | 9.2 |
| 0.9 | 13.7 |
| 1.0 | 24.0 |

**Table 7** CNO flux sensitivity (%) as a function of target mass, WbLS % and angular resolution for 5 years of data and 90% PMT coverage with the Borexino measured $^{40}$K level in water

| Target mass | WbLS | Angular resolution | | | |
|---|---|---|---|---|---|
| | | 25° | 35° | 45° | 55° |
| 50 kT | 0.5% | 14.5 | 20.9 | 26.6 | 31.5 |
| 50 kT | 1% | 13.8 | 20.1 | 25.7 | 31.0 |
| 50 kT | 2% | 13.7 | 20.2 | 25.8 | 31.2 |
| 50 kT | 3% | 12.4 | 17.7 | 22.6 | 27.1 |
| 50 kT | 4% | 11.8 | 16.8 | 21.3 | 25.6 |
| 50 kT | 5% | 11.4 | 16.1 | 20.5 | 24.6 |
| 25 kT | 0.5% | 20.0 | 29.2 | 36.9 | 45.5 |
| 25 kT | 1% | 19.3 | 27.3 | 34.6 | 42.4 |
| 25 kT | 2% | 17.9 | 25.9 | 32.6 | 39.0 |
| 25 kT | 3% | 17.1 | 24.2 | 30.8 | 36.5 |
| 25 kT | 4% | 16.3 | 23.2 | 29.3 | 35.3 |
| 25 kT | 5% | 15.6 | 22.2 | 28.2 | 33.9 |

**Table 8** Fit uncertainty for 5 years of data with the baseline detector configuration and the Borexino measured $^{40}$K level in water

| Signal | Normalization sensitivity (%) |
|---|---|
| $^8$B $\nu$ | 0.4 |
| $^7$Be $\nu$ | 0.9 |
| pep $\nu$ | 8.6 |
| CNO $\nu$ | 11 |
| $^{210}$Bi | 0.2 |
| $^{11}$C | 25 |
| $^{85}$Kr | 22 |
| $^{40}$K | 0.01 |
| $^{39}$Ar/$^{210}$Po | 45 |
| $^{238}$U chain | 0.02 |
| $^{232}$Th chain | 0.05 |

Due to the small fraction of scintillator in the WbLS, the relative $^{11}$C background is already reduced by an order of magnitude or more compared to pure scintillator experiments. Combined with directional reconstruction, the solar sensitivity becomes mostly insensitive to the $^{11}$C event rate at around the Borexino level, as shown in Table 9. This suggests that the overburden is not critical for a solar measurement with an experiment such as Theia.

**Table 9** Fit uncertainty on the CNO neutrino flux for 5 years of data with the baseline detector configuration as a function of isotope contamination level relative to baseline background level. For $^{210}$Bi this is the level of the out-of-equilibrium component in water relative to the out-of-equilibrium level in scintillator. For $^{85}$Kr and $^{39}$Ar this is the level of contamination in water relative to baseline

| Fraction of nominal contamination | Isotope | | | |
|---|---|---|---|---|
| | $^{11}$C | $^{210}$Bi | $^{85}$Kr | $^{39}$Ar |
| 0.1x | 5.3 | – | – | – |
| 1x | 5.3 | 5.3 | 5.3 | 5.3 |
| 10x | 5.4 | 5.4 | 5.3 | 5.3 |
| 100x | 5.5 | 6.0 | 5.4 | 5.4 |
| 1000x | – | 9.4 | 5.6 | 5.4 |
| 10000x | – | – | 5.9 | 5.5 |

**Table 10** Fit uncertainty for 5 years of data with the baseline detector configuration versus alpha and BiPo in-window and out of window rejection

| | CNO sensitivity (%) |
|---|---|
| Nominal | 5.3 |
| No $\alpha$ rejection | 5.4 |
| No BiPo in-window rejection | 5.4 |
| No $\alpha$, no BiPo in-window rejection | 5.4 |
| No $\alpha$, no BiPo rejection | 6.4 |

$^{210}$Pb, $^{210}$Bi, and $^{210}$Po are not necessarily in equilibrium with the rest of the uranium chain due to the long halflife of $^{210}$Pb and the possibility of Rn contamination. The level of out-of-equilibrium background achievable in ultrapure water has not been measured to the precision needed for this kind of experiment. The uranium chain contribution from the water component is already orders of magnitude larger than the non-equilibrium level measured in Borexino, so the non-equilibrium component in water must be many orders of magnitude larger than in scintillator as well to have any effect on the sensitivity. Table 9 shows that there begins to be a moderate impact when the out-of-equilibrium level in water is 1000× the out-of-equilibrium level in scintillator.

$^{85}$Kr and $^{39}$Ar are also found to have little impact on the sensitivity at up to 10,000× the expected level in scintillator, as shown in Table 9.

We find that alpha and Bi-Po rejection is not critical for this measurement. The largest effect comes from the ability to reject Bi-Po coincidences that occur in separate trigger windows, but there is little impact even assuming zero rejection, as shown in Table 10.

## 5 Conclusions

The feasibility of a low energy solar neutrino measurement with a large WbLS detector depends on the backgrounds and angular resolution achievable. At currently measured levels, the limiting background appears to be $^{40}$K in water. As this has not been a critical background in previous measurements, it is possible that lower levels may be achievable with more effort. The remaining unknown backgrounds in water have been shown to have little effect if they are kept to 1000–10,000 times the level achieved in scintillator. The $^{11}$C background is shown to be unimportant, which suggests that the overburden does not matter for this measurement. Finally, changes in scintillator and PMT coverage have been shown to have relatively small effects, which suggest that even by a 0.5% scintillator fraction, energy resolution is no longer the critical parameter to optimize. Instead, the impact of these changes on threshold, angular resolution, and energy systematic uncertainties will be the deciding factor. With a baseline detector of 50-kT total volume (50% fiducial), 90% PMT coverage and a 5% WbLS target, assuming a 25° angular resolution, a precision of several percent is possible for both CNO and pep neutrino fluxes.

Further studies hinge on additional R&D, including the Cherenkov detection efficiency enhancement provided by deployment of fast photon sensors [40–43], and demonstration of quenching and particle ID capabilities in WbLS. Development of a directional reconstruction algorithm would allow a direct demonstration of the required angular resolution discussed in this article. An exciting avenue for further exploration would be isotope loading of the WbLS target. Loading the WbLS target with an isotope such as $^7$Li for CC detection would provide an additional handle for signal/background separation via improved spectral information. Neutrinos interact in a pure (Wb)LS detector via ES. While the ES differential cross section is almost maximally broad, thus providing little handle on the underlying neutrino spectrum, that for the CC interaction on $^7$Li is extremely sharply peaked, resulting in the potential for high-precision measurement of the underlying neutrino energy spectrum. $^7$Li was proposed as an additive to a WCD in [44] for this reason; addition of this isotope to a WbLS detector would further improve the discrimination power. Initial studies are presented in [22]. A more quantitative study could yield significant improvements to the results presented here.

**Acknowledgements** The authors would like to thank the SNO+ collaboration for use of the solar neutrino generator, and scintillator optical model along with LAB/PPO properties. The authors would like to thank Wick Haxton for many useful discussions. This material is based upon work supported by the Director, Office of Science, of the U.S. Department of Energy under Contract No. DE-AC02-05CH11231 and by the U.S. Department of Energy, Office of Science, Office of Nuclear Physics, under Award Number DE-SC0010407.

to the original author(s) and the source, provide a link to the Creative Commons license, and indicate if changes were made.
Funded by SCOAP[3].

## References

1. Q.R. Ahmad et al. (SNO Collaboration), Phys. Rev. Lett. **87**, 071301 (2001)
2. Q.R. Ahmad et al. (SNO Collaboration), Phys. Rev. Lett. **89**, 011301 (2002)
3. K. Eguchi et al. (KamLAND Collaboration), Phys. Rev. Lett. **90**, 021802 (2003)
4. G. Bellini et al. (Borexino Collaboration), Phys. Rev. Lett. **107**, 141302 (2011)
5. Borexino Collaboration, Nature **512**, 383–386 (2014)
6. G. Bellini et al. (Borexino Collaboration), Phys. Rev. Lett. **108**, 051302 (2012)
7. J. Bahcall et al., Astrophys. J. **621**, L85 (2005)
8. N. Grevesse, J. Sauval, Space Sci. Rev. **85**, 161 (1998)
9. M. Asplund, N. Grevesse, J. Sauval, Nucl. Phys. A **777**, 1 (2006)
10. M. Asplund et al., Annu. Rev. Astron. Astrophys. **47**, 481 (2009)
11. A. Serenelli et al., Astrophys. J. **705**, L123 (2009)
12. A. Serenelli, W.C. Haxton, C. Peña-Garay, Astrophys. J. **743**, L24 (2011)
13. W.C. Haxton, A.M. Serenelli, Astrophys. J. **687**, 678 (2008)
14. S. Basu, ASP Conf. Ser. **416**, 193 (2009)
15. J.N. Bahcall, A.M. Serenelli, S. Basu, Astrophys. J. Suppl. **165**, 400–431 (2006)
16. F.L. Villante et al., Phys. Lett. B **701**, 336–341 (2011)
17. D.G. Cerdeno, J.H. Davis, M. Fairbairn, A.C. Vincent, (2017). arXiv:1712.06522
18. M. Yeh et al., Nucl. Instrum. Methods Phys. Res. A **660**, 51–56 (2011)
19. M. Wurm et al., Astropart. Phys. **35**, 685–732 (2012)
20. C. Buck, M. Yeh, J. Phys. G **43**, 093001 (2016)
21. G. Orebi Gann, (2015). arXiv:1504.08284 [physics.ins-det]
22. J.R. Alonso et al., (2014). arXiv:1409.5864v3 [physics.ins-det]
23. C. Aberle, A. Elagin, H.J. Frisch, M. Wetstein, L. Winslow, J. Instrum. **9**, P06012 (2014)
24. A. Elagin et al., Nucl. Instrum. Methods A **849**, 102 (2017)
25. J. Caravaca et al., Phys. Rev. C **95**, 055801 (2017)
26. J. Caravaca et al., Eur. Phys. J. C **77**, 811 (2017)
27. M. Sakai, PhD thesis, Hawaii U. (2016)
28. J. Kumar, J.G. Learned, M. Sakai, S. Smith, Phys. Rev. D **84**, 036007 (2011)
29. R. Bonventre et al., Phys. Rev. D **88**, 053010 (2013)
30. B. Aharmim et al. (SNO Collaboration), Phys. Rev. C **75**, 045502 (2007)
31. K. Abe et al. (Super-Kamiokande Collaboration), Phys. Rev. D **83**, 052010 (2011)
32. H. O'Keeffe, E. O'Sullivan, M. Chen, Nucl. Instrum. Methods Phys. Res. A **640**, 119–122 (2011)
33. J. Brack et al., Nucl. Instrum. Methods Phys. Res. A **712**, 162–173 (2013)
34. M. Balata et al., Nucl. Instrum. Methods Phys. Res. A **370**, 605–608 (1996)
35. C. Arpesella et al. (Borexino Collaboration), Phys. Rev. Lett. **101**, 091302 (2008)
36. G. Alimonti et al. (Borexino Collaboration), Nucl. Instrum. Methods Phys. Res. A **609**, 58–78 (2009)
37. H. O'Keeffe, Ph.D. thesis, Oxford University (2008)
38. A.J. Noble et al., Scientific Review of SNO Water Systems. SNO+ Document. 1870v1 (1996)
39. S.W. Li, J.F. Beacom, Phys. Rev. C **89**, 045801 (2014)
40. B. Adams, A. Elagin, H. Frisch, R. Obaid, E. Oberla, A. Vostrikov, R. Wagner, M. Wetstein, Nucl. Instrum. Methods A **732**, 392 (2013)
41. B. Adams et al., Rev. Sci. Instrum. **84**, 061301 (2013)
42. B.W. Adams et al. (LAPPD Collaboration), JINST (2016). arXiv:1603.01843 [physics.ins-det]
43. O.H.W. Siegmund, J.B. McPhate, J.V. Vallerga, A.S. Tremsin, H.E. Frisch, J.W. Elam, A.U. Mane, R.G. Wagner, J. Instrum. **9**, C04002 (2014)
44. W.C. Haxton, Phys. Rev. Lett. **76**, 1562 (1996)